# ENABLING ROBUST SENSOR NETWORK DESIGN WITH DATA PROCESSING AND OPTIMIZATION MAKING USE OF LOCAL BEEHIVE IMAGE AND VIDEO FILES

Ephrance Eunice Namugenyi (PhD)[1] David Tugume (MSc)[2] Augustine Kigwana (BSc)[3] and Benjamin Rukundo (BSc)[4]

[1]Department of Computer Networks, CoCIS, Makerere University, Uganda
AdEMNEA Project

***ABSTRACT***

*There is an immediate need for creative ways to improve resource efficiency given the dynamic nature of robust sensor networks and their increasing reliance on data-driven approaches. One key challenge faced is efficiently managing large data files collected from sensor networks for example optimal beehive image and video data files. We offer a revolutionary paradigm that uses cutting-edge edge computing techniques to optimize data transmission and storage in order to meet this problem. Our approach encompasses data compression for images and videos, coupled with a data aggregation technique for numerical data. Specifically, we propose a novel compression algorithm that performs better than the traditional Bzip2, in terms of data compression ratio and throughput. We also designed as an addition a data aggregation algorithm that basically performs very well by reducing on the time to process the overhead of individual data packets there by reducing on the network traffic. A key aspect of our approach is its ability to operate in resource-constrained environments, such as that typically found in a local beehive farm application from where we obtained various datasets. To achieve this, we carefully explore key parameters such as throughput, delay tolerance, compression rate, and data retransmission. This ensures that our approach can meet the unique requirements of robust network management while minimizing the impact on resources. Overall, our study presents and majorly focuses on a holistic solution for optimizing data transmission and processing across robust sensor networks for specifically local beehive image and video data files. Our approach has the potential to significantly improve the efficiency and effectiveness of robust sensor network management, thereby supporting sustainable practices in various IoT applications such as in Bee Hive Data Management.*

***KEYWORDS***

*Data Processing and Optimization, Edge Computing, Robust Sensor Networks, Data Compression, Data Aggregation, Image and Video Data Files*

## 1. INTRODUCTION: BACKGROUND, OBJECTIVES AND PAPER CONTRIBUTIONS

### 1.1. Background

Sensor network-based applications, like bee hive management [1], encounter increasing difficulties, especially when managing large data files gathered via sensor networks. These materials provide difficult challenges for storage and transportation, particularly in environments with limited resources. With the use of technologies like sensor networks, data-driven methodologies have transformed system operations and come with pros and cons for different applications. Local practitioners struggle with the complex work of handling sensor data, which is made more difficult by the limits of existing technologies like GSM and Wi-Fi [2]. This is especially true in underdeveloped countries like Uganda. These limitations impede the development of Internet of Things (IoT) applications. A new technology that arises to address these issues is edge computing [3, 4].

Edge computing minimizes latency and improves bandwidth efficiency by relocating data storage and computation closer to the network's edge, where data is generated and consumed. Edge-based methods for data aggregation and reduction [5] in audio and video provide a way to reduce files without sacrificing quality. As a result, data transmission and storage efficiency are increased, which is important in places with limited resources. Edge computing [1, 3] has the transformative potential to improve decision-making and promote efficiency across diverse IoT applications by enabling practitioners to collect and analyze more extensive data. By monitoring environmental indicators like temperature, humidity, and activity, practitioners can use edge computing to detect abnormalities and problems early on. Furthermore, edge computing enables practitioners to create predictive models for foreseeing and proactively addressing future difficulties and maximizing resource use through real-time data insights. Although edge computing is still being adopted in IoT applications, there are several potential advantages [6]. By streamlining data storage and transmission, edge computing can help overcome resource-constrained environments' obstacles and open the door to data-driven precision in a variety of applications.

## 1.2. Objective

This paper presents an edge computing-based solution to the complex challenges faced in transferring large data files across robust sensor networks using WiFi and GSM [2]. We propose a sophisticated compression algorithm for images and videos, as well as a data aggregation technique for numerical data. These advancements aim to improve data transfer efficiency and support sustainable and resource-efficient robust sensor network management when developing various IoT applications.

## 1.3. Paper Contributions

***Technical contributions:*** The study proposes a compression technique designed to handle the unique properties of beehive data, which includes photos, and videos. This approach outperforms classical techniques in terms of compression ratio, which is especially useful in settings with limited resources. The study as a bonus includes a data aggregation approach, which are intended to improve the accuracy of machine learning models trained on beehive numerical data, in addition to the compression algorithm. This enhancement makes it easier to comprehend hive dynamics given our choice data, giving users more accurate information to use when making decisions for better management.

***Research contributions:*** The study does a thorough evaluation of different compression methods in both Wi-Fi and GSM networks, and the results show notable gains in transmission time, especially in low-bandwidth situations. Through a comparative analysis of their flexibility in various network situations, the research provides insightful information for future use, helping to determine the best compression techniques. It highlights the significance of taking network features into account when optimizing data transfer for resilient sensor networks and investigates the implications of compression approaches for energy efficiency, which is critical for battery-powered devices. The research sets the way for future studies in sensor networks and data transmission optimization, with applications extending to different contexts including bee-hive data management. It suggests segmentation, compression, and aggregation to improve big file transfers.

## 2. ROBUST SENSOR NETWORK DESIGN & DATA TRANSFER CHALLENGES

Effective data transport is crucial for modern Internet of Things applications [1]. However, a variety of obstacles, such as resource limitations, excessive latency, and bandwidth limitations, face practitioners when attempting to manage sensor networks seamlessly [7]. These difficulties are especially noticeable in underserved and distant places where IoT devices are used. IoT applications confront a variety of difficulties, such as resource limitations, excessive latency, and constrained bandwidth. Sensor networks produce a variety of data kinds, including as pictures, videos, and numerical data, which makes sending them over networks with limited capacity an expensive and time-consuming process that prevents practitioners from getting real-time insights. For real-time IoT applications, high latency—the delay in data transmission—emerges as a major obstacle that affects crucial operations like early anomaly detection. Furthermore, practitioners struggle to invest in the necessary infrastructure for reliable data transfer capabilities due to resource constraints. The intricacies that Internet of Things practitioners confront in guaranteeing smooth and effective data transport in contexts with limited resources are highlighted by these coupled difficulties. The effectiveness of IoT activities is greatly impacted by these difficulties. The lack of access to real-time data makes it more difficult for practitioners to recognize and resolve problems in a timely manner, which can lead to operational inefficiencies and financial losses.

Wi-Fi sensor network topologies that are enhanced by edge computing show promise [3, 4]. Edge computing overcomes the drawbacks of traditional cloud-based solutions by decreasing latency and boosting bandwidth efficiency; this is especially helpful in environments with limited resources. Techniques for data aggregation and compression are additional cutting-edge technologies that allow for the effective transfer of data files without sacrificing quality [7]. Furthermore, dependable and energy-efficient substitutes are provided by low-power wireless technologies as LoRaWAN [9] and satellite-based data transmission [8]. By offering practitioners real-time information to improve the effectiveness and efficiency of their operations inside reliable sensor network architectures, these cutting-edge solutions have the ability to completely transform data transfer in a variety of Internet of Things applications.

## 3. DATA OPTIMIZATION TECHNIQUES USING EDGE COMPUTING: COMPRESSION AND AGGREGATION

Compression Algorithm Design

Robust data sensor networks confront a myriad of data transfer challenges, exacerbated by limited bandwidth, high latency, and resource constraints. In response to these intricate challenges, this section unveils a novel compression algorithm meticulously designed for images and videos, specifically tailored to the needs of beehive data management. This groundbreaking algorithm performs much better than traditional methods, such as the Bzip2 [5,10] compression algorithm, revealing substantially improved compression ratios. This marks a significant leap forward in optimizing the efficiency of data transfer in the realm of beehive management, enabling beekeepers to transmit and store critical data with greater ease, affordability, and scalability.

**Key Features of the Proposed Compression Algorithm [5]**

- Adaptive compression: The algorithm dynamically adapts to the unique characteristics of beehive data, including image and video formats, to achieve optimal compression ratios.
- Lossless compression: The algorithm compresses data without sacrificing quality, ensuring that beekeepers retain access to accurate and reliable data insights.
- Fast compression and decompression: The algorithm is designed to achieve high compression and decompression speeds, even on resource-constrained devices.

*Average Compression Ratio (ACR) = Σ (Original File Size) / Σ (Compressed File Size)*
*Bit-Plane Error Rate (BPER) = Σ (Number of Incorrect Bits) / (Total Number of Bits)*
*Compression Time (CT) = Time to Compress Data*
*Decompression Time (DT) = Time to Decompress Data*

Data Aggregation for Numerical Data

In tandem with advancements in image and video compression, this subsection delves into the application of data aggregation techniques for textual data as an additional bonus to the paper, specifically CSV files containing critical parameters like temperature, humidity, and $CO_2$. In aggregate multiple smaller data packets into larger ones before transmission to reduce overhead and improve efficiency. This is particularly useful for sensor networks where multiple sensors can contribute data for a single transmission. And this is aimed at reducing the overhead to reduce network traffic in a medium.

**Key features of Data Aggregation Algorithm**

- Accuracy: The method preserves the original data's accuracy. This implies that there aren't any appreciable inaccuracies and that the combined data accurately reflects the original data.
- Efficiency: Both in terms of memory use and computational complexity, the algorithm is very efficient. For big numerical datasets where performance is crucial, this is significant.
- Scalability: The algorithm is able to adjust to changes in the distribution of the data and is scalable enough to handle big datasets. This holds significance for real-time applications and continuously expanding databases.

### 3.1. Conclusion

The innovative compression and aggregation techniques [11] presented in this chapter have the potential to revolutionize data transfer and analysis in robust sensor network management. By optimizing the efficiency and accuracy of data transmission, these techniques can help users of various applications to improve the functionalities of their networks.

## 4. RELATED WORKS

Our study expands upon a wealth of prior research in edge computing and data optimization methodologies specifically using local bee-hive video and image data files.. In this chapter, we provide a comprehensive review of the related literature, highlighting the current state of knowledge and identifying gaps. This review informs our unique approach to data optimization of large data files across robust sensor networks in local beehive management.

Data optimization strategies have been the subject of extensive research, with the goal of minimizing data file sizes without compromising quality. Common techniques include compression, aggregation, and sampling. Compression algorithms [10] reduce the size of data files by exploiting redundancy and statistical properties of the data. Aggregation techniques combine similar data points into a single representative value, thereby reducing the amount of data that needs to be transmitted or stored. In order to lessen the computing load of data processing and analysis, sampling algorithms choose a subset of data points from a larger dataset. However, there is limited research on the application of edge computing [3,4] to data optimization in various robust sensor networks for resource-constrained environments. Particularly, there is a scarcity of research addressing the utilization of edge computing for the development of data optimization methods tailored to the unique characteristics of beehive data, and the evaluation of the effectiveness of these optimization methods in practical robust network management scenarios [12, 13, 14, 15]. Closing these gaps is crucial for advancing our understanding and implementation of efficient data optimization strategies in the context of beekeeping.

In related works, the Implementation of IoT solutions for beekeeping applications has been somewhat restricted. Even while electronic remote beehive monitoring and the Internet of Things are well-established, very few applications combine these sectors. For instance, Lyu et al. described a smart beehive system in [16] that uses the General Packet Radio Service (GPRS) network to track temperature, humidity, weight, attitude, and GPS location. After that, the beehive sends its data to a monitoring center, where a worker can examine it and ascertain the hive's condition. The operator can notify the beekeeper in the event of anomalies. This strategy, however, can have drawbacks because it depends on the monitoring center having an intermediary on staff. Creating a dashboard with an automated warning system integrated and direct data access is required to solve this problem [17]. IOHIVE, an IoT-based platform that assists beekeepers in monitoring the temperature, humidity, and weight of the hive, is presented by Chamaidi et al. in [18]. A beehive monitoring system based on IoT and microservices, BHiveSense was most recently introduced by Cota et al. [19]. It consists of a number of components, including a web application, a mobile application, a REST back-end API, and a low-cost hive-sensing prototype. By tackling interoperability, scalability, agility, and maintenance challenges, the authors want to improve the sustainability and integration of beekeeping activities and ultimately provide an effective beehive monitoring system.

Additionally, certain systems provide the use of video surveillance (Meitalovs et al., 2009). In order to follow several viral infections, including V. destructor mites, Chen et al. (2020) study thermal pictures, although they have not yet produced any findings. To track the extent of V. destructor infestation, Bjerge et al. (2019) examine video sequences that have been recorded. While testing several camera configurations for the purpose of visually identifying infested bees, Schurischuster et al. (2016) did not suggest an algorithm for mite detection; instead, they concentrated solely on capturing high-quality footage. Schurischuster et al. (2018) use machine learning and image analysis approaches to identify individual bee photos as either mite-infested or not. A method for identifying V. destructor mites inside honeybee cells is put forth by Elizondo et al. (2013). Nevertheless, the system isn't perfect and cannot collect image data.

Our unique study approach therefore fills in the gaps in the literature by creating a novel framework for robust sensor network management through data optimization. By leveraging edge computing, our architecture lowers latency and increases bandwidth efficiency by bringing in data optimization techniques. Furthermore, data aggregation and compression methods specifically designed for beehive data are integrated into our architecture. We evaluate the performance of our framework in a real-world beehive management scenario [1] and demonstrate that it can significantly reduce the size of data files without sacrificing quality. This can lead to significant cost savings and improved efficiency for the network that can be applied in various similar applications.

In conclusion, our research on data optimization of large data files across robust sensor networks in using local beehive management data is a significant contribution to the field. Our approach has the potential to transform data processing and transmission techniques by filling in the gaps in the current literature and creating a fresh framework for data optimization.

## 5. Sensor Network Design and Conceptual Framework

Network Architecture

Building a sensor network for beehive management is difficult because of data transit issues in agricultural environments, particularly low bandwidth in remote areas and inadequate network coverage. In order to overcome these obstacles, we suggest a sensor network design that combines GSM and Wi-Fi [7]. While GSM provides dependable coverage in underserved areas, Wi-Fi allows for quick data transfer over short distances. The parts consist of beehive nodes that gather data from the sensors, send it via GSM or WiFi that act as our major communication modules, then to the cloud server and finally to the application. Depending on deployment requirements, this flexible design supports many communication modules, including Wi-Fi, Lo-Ra SX1278 [8], and SIM800L EVB [7], with Wi-Fi and the GSM modules being prioritized initially for functional demonstration and prototyping.

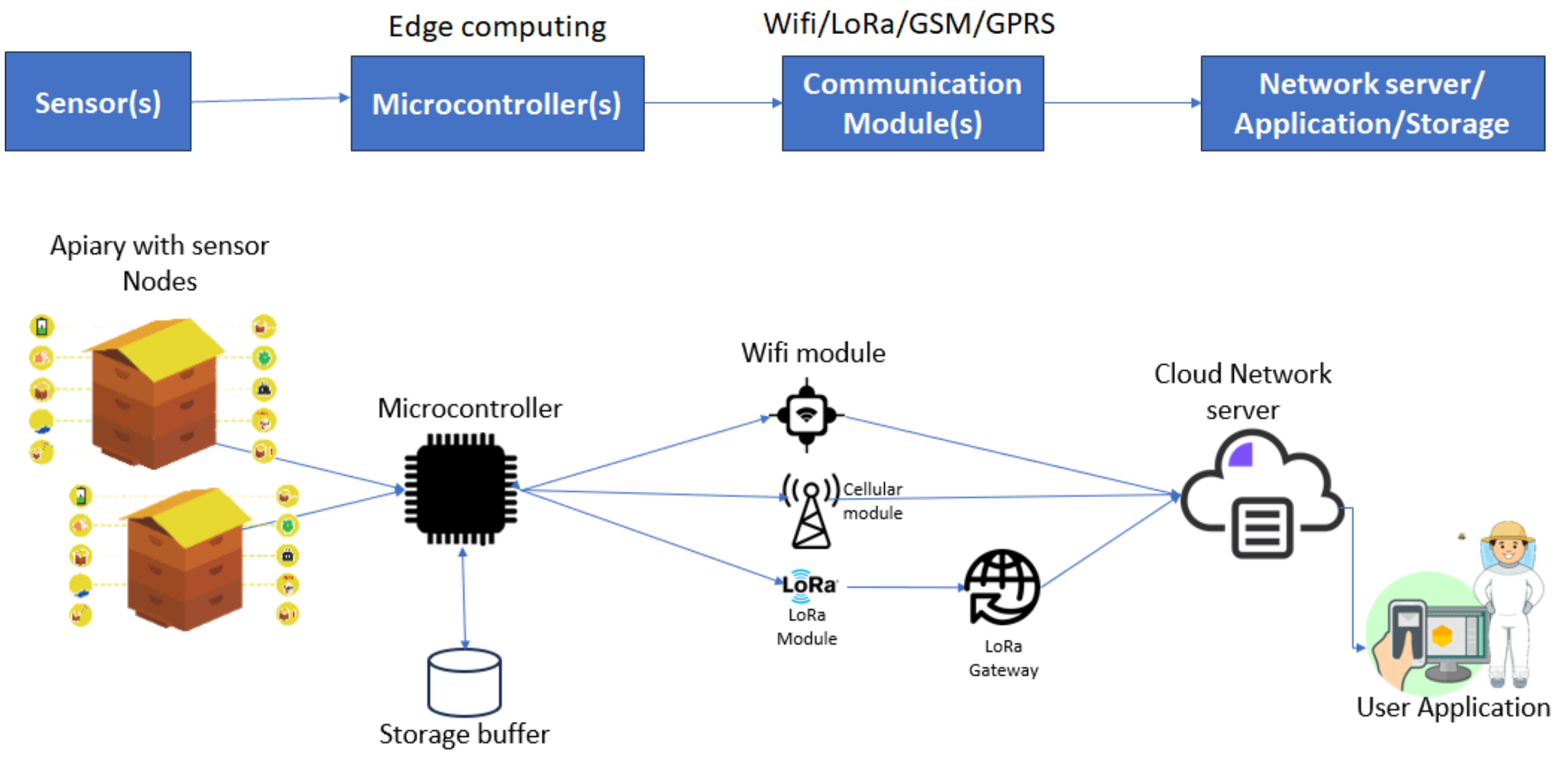

Figure 1:Network Design Architecture and Flow Diagram for Local Bee-hive Management Case

A Raspberry Pi microcontroller, which connects to sensors for localized processing and minimizes latency and bandwidth requirements, is at the heart of our system. JPEG and PNG compression methods, in particular, maximize the efficiency of data transmission when dealing with images. Numerical data aggregation techniques like summing and averaging combine points and minimize the need for separate transmissions. Based on deployment criteria, modules such as Wi-Fi, Lo-Ra SX1278, and SIM800L EVB can be selected for the transmission of processed data, with Wi-Fi and GSM being used for prototyping initially. In order to maximize bandwidth, the chosen module establishes connectivity while drastically reducing delivered data. Analysis and decision-making in real-time or almost real-time are made possible by edge aggregation and compression, which reduce latency.

In summary, our network architecture design leverages the capabilities of edge computing, with an initial emphasis on Wi-Fi and GSM, using a Raspberry Pi microcontroller for local bee-hive sensor data processing. Through the synergistic application of compression and aggregation techniques, our architecture enables the transmission of concise video, image and numerical data segments, aligning with the requirements of any robust sensor network application. The chosen communication module establishes an efficient bridge to the cloud or network server, supporting real-time insights into beehive ecosystems while efficiently utilizing network resources. Studies with LoRa module design are ongoing.

### Key Parameters

The robust sensor network design for beehive data requires a number of key parameters, such as *throughput*, which measures the amount of data sent over the network in a given amount of time; *delay tolerance*, which indicates the maximum amount of time before data delivery is necessary and is essential for real-time monitoring in any IoT application; *compression rate*, which measures the reduction of data size through the use of compression algorithms to minimize bandwidth requirements; *data retransmission*, which ensures reliable delivery in areas with inadequate network coverage; *bandwidth optimization*, which applies techniques to improve data transfer efficiency and lower costs; and *time* taken, which is the amount of time needed to gather, send, and process data that for example beekeepers in our specific case need to know in order to receive timely insights on the health of their hives.

### Conceptual Framework for Beehive Data Management

Our theory of beehive data management uses sensor nodes that are positioned strategically and have a variety of sensors attached to them in order to gather important information. By utilizing Raspberry Pi microcontrollers for interface, these nodes facilitate edge computing for data integration. The Raspberry Pi performs data transformation tasks by using our suggested compression algorithm for encoding and decoding image (JPEG) and video files. The aggregation algorithm works on numerical data compression. The main communication modules are Wi-Fi and the GSM SIM800L EVB, which can be modified for use in upcoming experiments with modules such as Lo-Ra SX1278. The Raspberry Pi can be connected to a network server or the cloud via communication modules, enabling data-driven insights and real-time monitoring. Applications for tasks like historical data analysis and well-informed decision-making are hosted on the server. Dynamic modifications are made possible by a feedback loop that links the beehive environment and the server, guaranteeing real-time adaptability. The framework ensures effective file transfer specifically for beehive data management and accommodates potential LoRa integration for resource-constrained applications (experiments ongoing). It is optimized for lower bandwidth requirements and incorporates Wi-Fi and GSM modules.

```
Sensors -> Sensor Nodes -> Raspberry Pi -> Cloud/Network Server
                               |                    |
                               |                    |
                      Data Transformation     Data Transfer
                               |                    |
```

```
       |                  |
Image Compression    Wi-Fi, GSM,
       |                  |
       |                  |
Numerical Compression   Feedback Loop
```

Conclusion

The design of a robust sensor network and conceptual framework for data optimization are essential for enabling beekeepers in our specific study case scenario to collect, transmit, and analyze large data files in real-time. By addressing the unique challenges of beehive data management, our proposed approach can help beekeepers to improve hive health, increase honey production, and reduce costs.

# 6. ALGORITHM DESIGN AND FUNCTIONALITY

## 6.1. Introduction

This section provides an exhaustive breakdown of the proposed compression algorithm for images and videos [13, 14]. It elucidates the superior performance of our proposed algorithm compared to conventional Bzip2, showcasing its efficiency in optimizing data transfer for beehive management. Furthermore, the data aggregation algorithm's utility for numerical data is examined, with a focus on how it might improve data transmission speed and accuracy.

## 6.2. Suggested Algorithm for Video Compression

A format and technique for video compression is called "video coding," which transforms digital video into a form that can be broadcast or stored with less space usage. An encoder converts a video into a compressed format and a decoder converts the video back into uncompressed format.

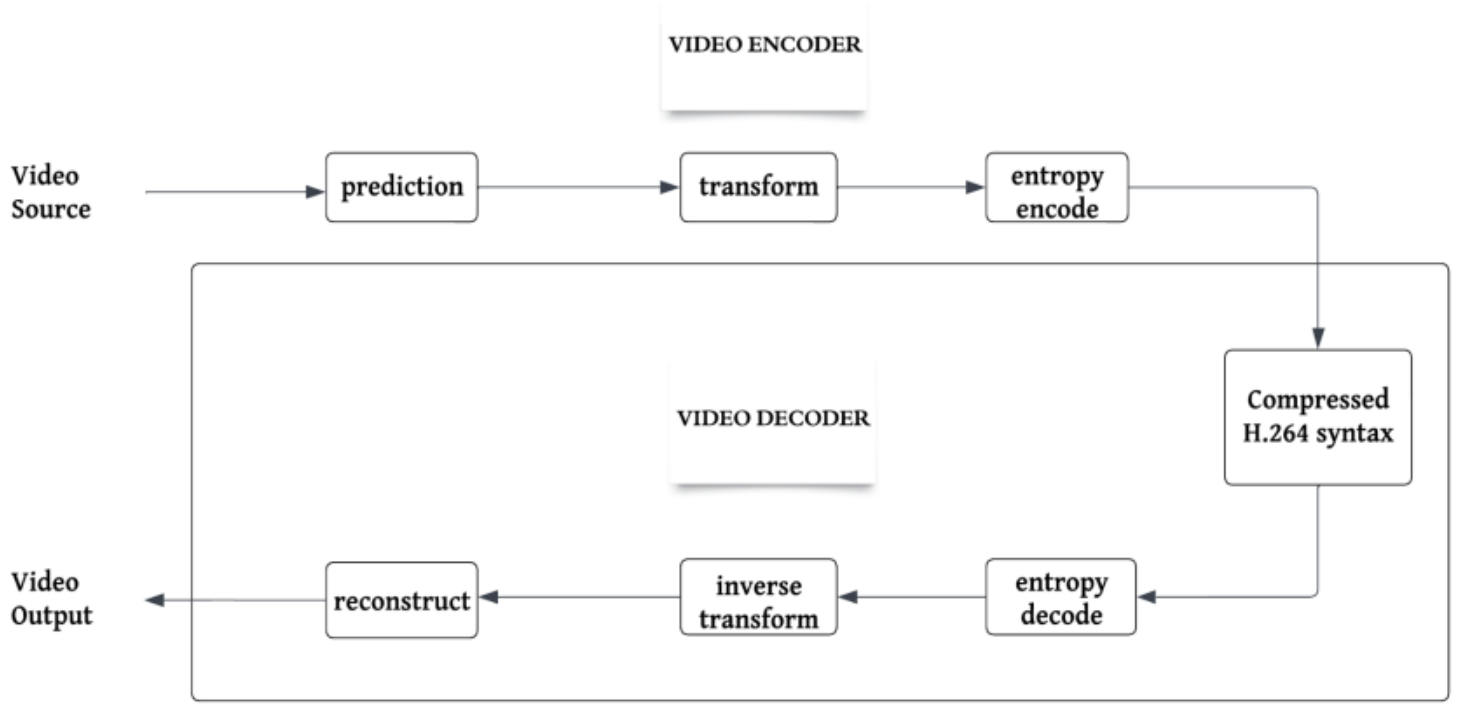

Figure 2: Algorithm coding and Decoding Process

A compressed bitstream is produced by the encoder after it has predicted, transformed, and encoded (Figure 2). The decoder, on the other hand, produces a decoded video sequence by performing decoding, inverse transformation, and reconstruction. A sequence of original video frames or fields is encoded using the specified algorithm format (Figure 3), creating a compressed representation in bits. Decoding the compressed bitstream enables storage or transport, and reconstructing the original video. However, because some image quality is lost during compression, the decoded version typically looks different from the original sequence.

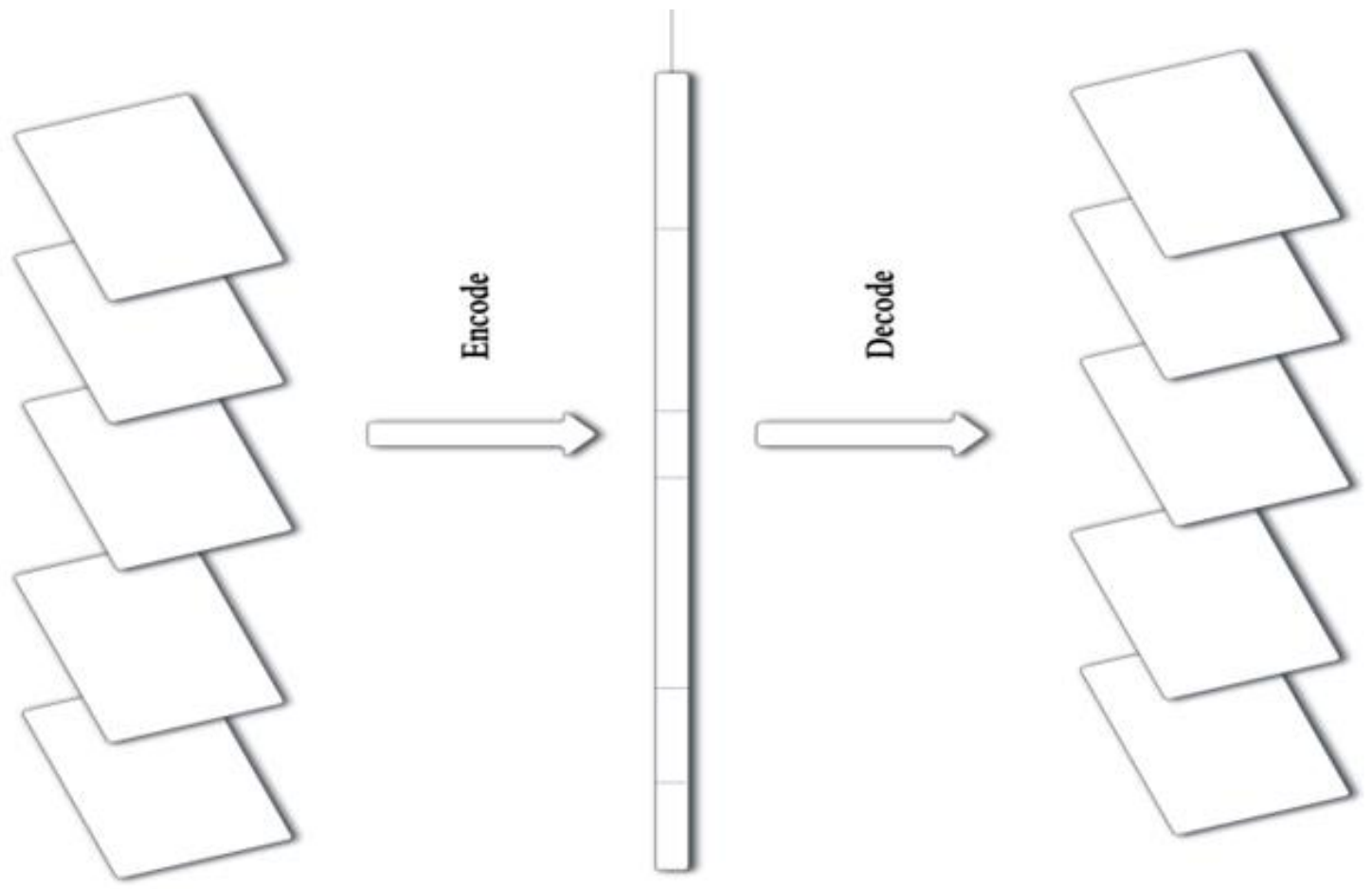

Figure 3: Video Coding: Source Frames, Bitstream Encoded, and Frames Decoded

Suggested Algorithm for Image Compression

The suggested lossy compression technique produces noticeably reduced file sizes with negligible to no effect on the sharpness and quality of the images. An original image can be 10 times smaller when it is compressed into a.jpeg file. The way this algorithm operates is that it retains the information that the human eye is capable of seeing while eliminating information that is difficult for it to see. This is a condensed description of the algorithm's operation:

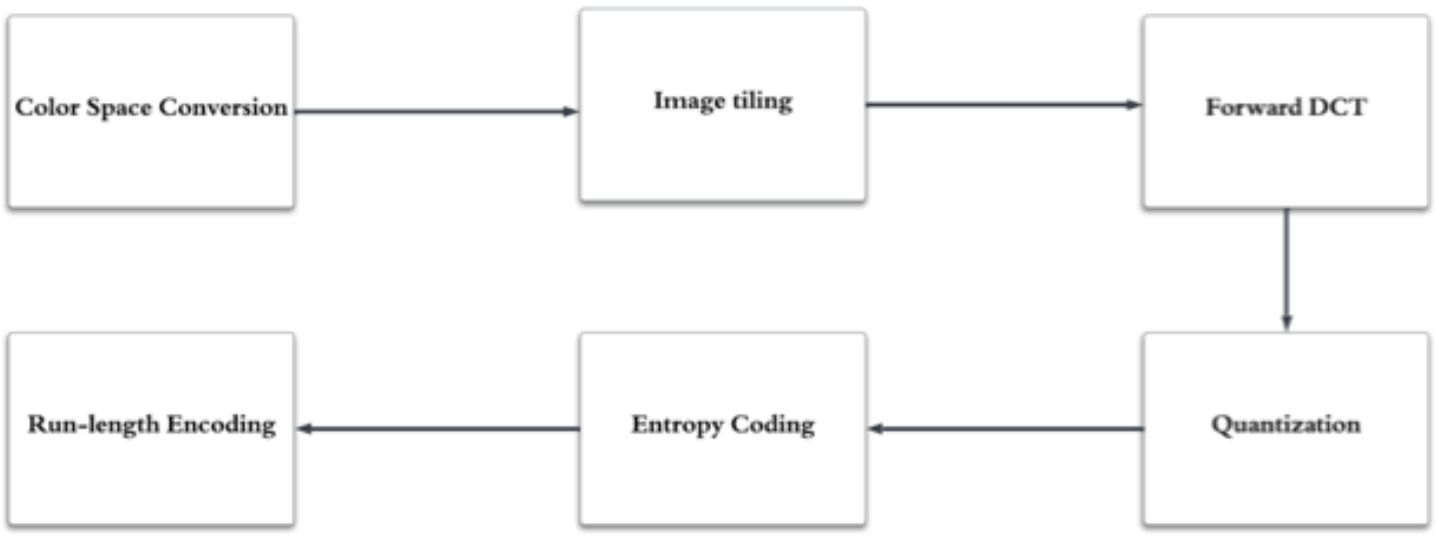

Figure 4: Major Steps involved in the proposed image compression algorithm

In order to prioritize brightness sensitivity above color perception, the original RGB color space is converted to YCbCr at the start of the image compression process. The picture is then separated into 8x8 pixel segments for separate processing. For effective representation, each block is subjected to the Discrete Cosine Transform (DCT), which converts pixel values into frequency domain coefficients. The next step is quantization, which divides the DCT coefficients according to predetermined matrices. Aggressive quantization of the high-frequency components causes data loss. By assigning longer codes to infrequent values, Huffman coding maximizes compression by encoding quantized coefficients. By effectively encoding repeated values, run-length encoding significantly minimizes the amount of the data. Next, the file header and metadata information are added to the compressed data bundle. The steps are reversed during decoding: file parsing reads compression settings, inverse quantization reverses quantization, Huffman decoding returns quantized DCT coefficients, and inverse DCT recreates 8x8 blocks. When YCbCr data is converted to RGB using color space conversion, the decompressed image is prepared for display or additional processing. It's important to remember that lossy compression, which has customizable compression levels from 1 to 100 that indicate different degrees of quality and file size, compromises some image quality in exchange for reduced file sizes.

Aggregation Algorithm For Numerical Data

The data aggregation algorithm as a bonus is designed to efficiently collect and store sensor data, including temperature, humidity, carbon dioxide levels, and weight and only initiate transmission to the server only when the file size reaches a predetermined threshold. The key features of this algorithm include

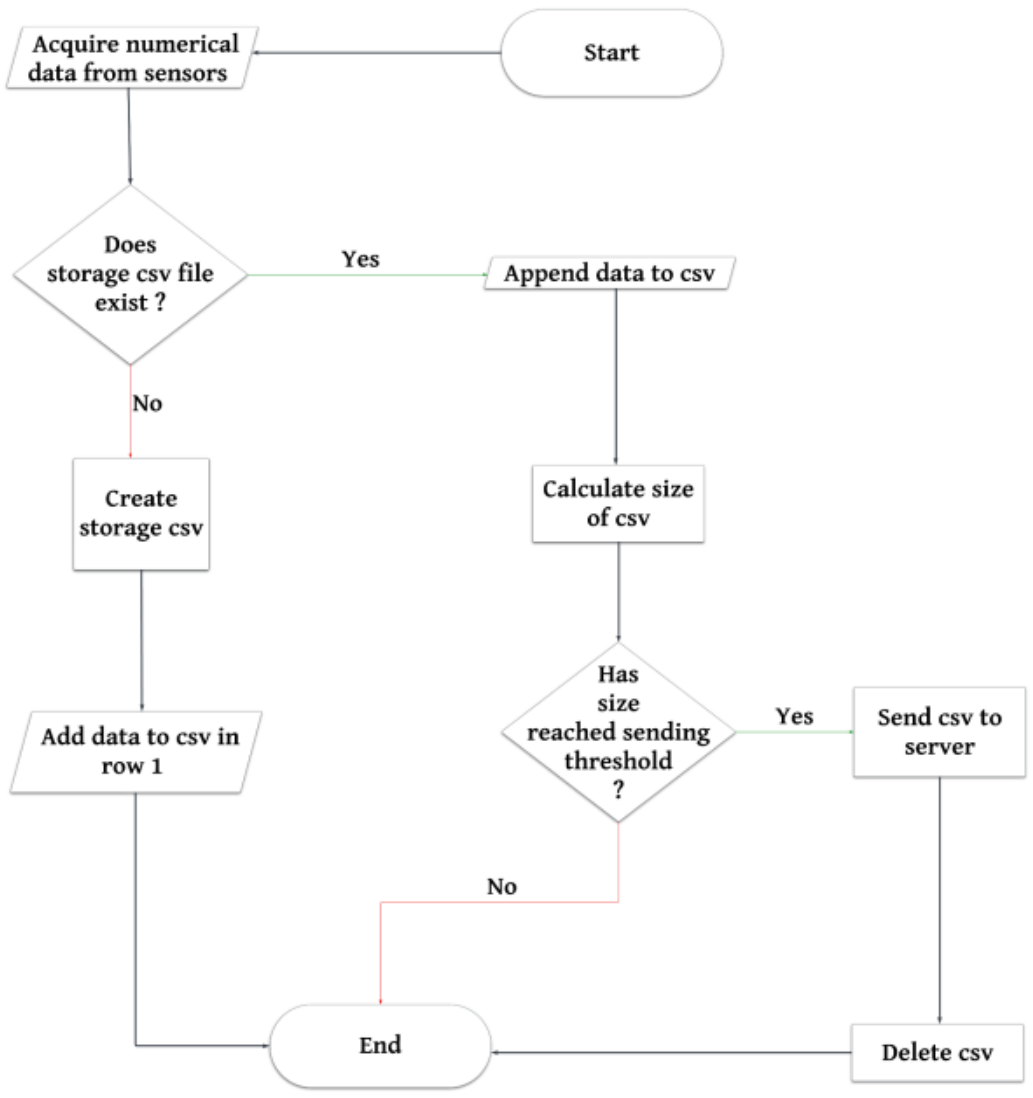

Figure 5: The data aggregation algorithm

First, sensor data must be gathered and prepared. CSV files are then used to manage the data. The method adds new data to an existing file or, if necessary, creates a new one after determining whether a given name is present in a CSV file. The process of continuous data appending guarantees that every data point is kept independently within a single CSV file. Additionally, the system keeps track of the file size and only transmits data when it crosses a certain threshold. This method reduces data transmission overhead and maximizes bandwidth use. Ultimately, the algorithm starts the process of sending data to the server when the file size satisfies the requirements for transmission.

## 7. RESULTS AND ANALYSIS

### 7.1. Introduction:

This section embarks on a comprehensive analysis of experimental results, showcasing the superior performance of our algorithm in terms of data compression rate and time efficiency. Statistical analyses and visual representations substantiate our findings, providing an in-depth understanding of the algorithm's prowess.

Analysis and Evaluation of Video Compression Algorithm Experiments

Compression algorithms play a pivotal role in optimizing data transmission, especially in scenarios where limited bandwidth and high data volumes are significant challenges. In this analysis, we delve into the results of the experiments conducted on our chosen compression algorithms in comparison with bzip2. The experiments involved compressing a 135MB video, among others, and a 24MB image along with different other sizes to evaluate the effectiveness of these algorithms.

Compression Ratios and File Size Reduction:

The primary goal of compression algorithms is to reduce file sizes while retaining essential data. The results reveal that both choice algorithms achieved substantial reduction in file sizes for both the video and image [13, 14]. For the video, the original 135MB file was reduced to 5.5MB, while bzip2 reduced it to 6.8 MB. Similarly, for the image, the choice algorithm and bzip2 compressed the 24MB file to 10.3MB and 11.2MB, respectively. This showcases the effectiveness of both algorithms in achieving significant file size reduction.

Table 1. Compression Ratio vs File Size Reduction

| **Raspberry pi camera module v3** | | | | Camera Pixel Setting 1280x720 | |
|---|---|---|---|---|---|
| | | frame rate 8 | | | |
| **Length of Video(s)** | **Original Size(bytes)** | **Bzip2 Final Size(bytes)** | **Video Algorithm Final Size(bytes)** | **Bzip2 Compression Ratio** | **Video Algorithm Compression Ratio** |
| 2 | 14,929,920 | 9,852,909 | 134,101 | 1.52 | 111.33 |
| 4 | 41,472,000 | 27,280,319 | 373,718 | 1.52 | 110.97 |
| 6 | 68,014,080 | 44,698,342 | 601,929 | 1.52 | 112.99 |
| 8 | 94,556,160 | 62,090,322 | 752,216 | 1.52 | 125.70 |
| 10 | 121,098,240 | 79,464,023 | 972,339 | 1.52 | 124.54 |
| 12 | 147,640,320 | 96,879,797 | 1,179,606 | 1.52 | 125.16 |
| 14 | 159,252,480 | 104,461,325 | 1,364,544 | 1.52 | 116.71 |
| 16 | 200,724,480 | 131,728,216 | 1,595,435 | 1.52 | 125.81 |
| 18 | 225,607,680 | 148,077,251 | 1,907,795 | 1.52 | 118.26 |
| 20 | 253,808,640 | 166,534,384 | 2,332,089 | 1.52 | 108.83 |

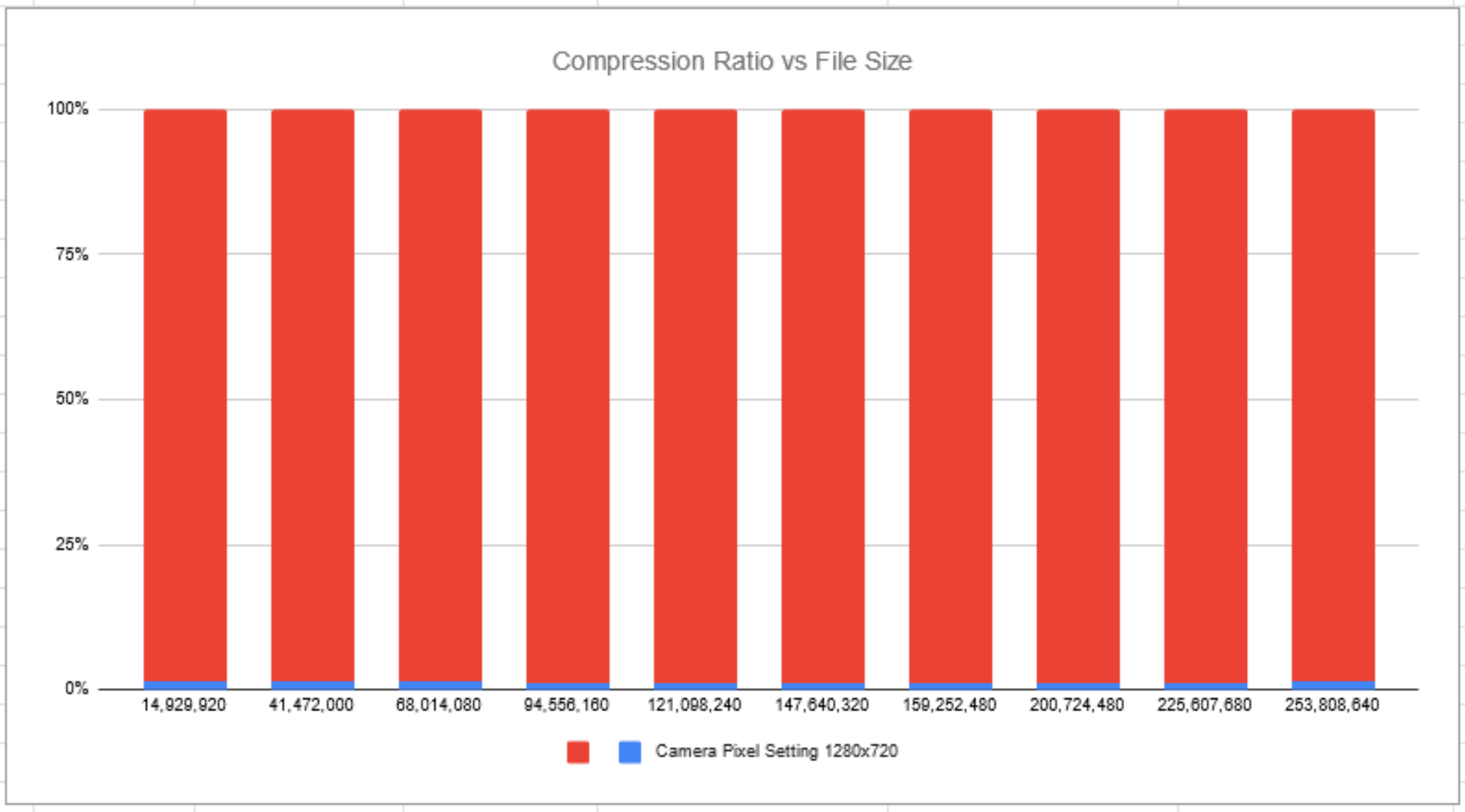

Figure 7: Compression Ratio Vs File Size

Table 2. Shows experiment results for transmitting the video compressed files using Wi-Fi and cellular communication technologies.

| **Video Algorithm Final Size(bytes)** | **Average sending time Wi-Fi(s)** | **Avg Bytes Sent Per Sec Wi-Fi** | **Average sending time GSM (s)** | **Avg Bytes Sent Per Sec GSM (s)** |
|---:|---:|---:|---:|---:|
| 134,101 | 1.567 | 88,060.40 | 209.03 | 1592.10 |
| 373,718 | 1.700 | 224,299.03 | 584.97 | 649.57 |
| 601,929 | 1.767 | 343,393.97 | 461.10 | 1599.70 |
| 752,216 | 1.900 | 401,925.00 | 1680.00 | 604.17 |
| 972,339 | 1.833 | 523,617.50 | 1238.93 | 917.03 |
| 1,179,606 | 1.967 | 599,785.93 | 1306.27 | 1056.67 |
| 1,364,544 | 1.833 | 740,028.13 | 1655.57 | 1227.87 |
| 1,595,435 | 2.100 | 769,461.83 | 1594.63 | 1149.80 |
| 1,907,795 | 1.900 | 1,015,327.33 | 2833.47 | 731.00 |
| 2,332,089 | 1.967 | 1,185,883.50 | 2489.97 | 963.53 |

Video Compression and Performance Analysis

To analyze the performance improvement achieved by different compression algorithms when transferring large files across both Wi-Fi and GSM/GPRS networks, we'll compare the transmission times before and after compression and calculate the percentage decrease in transmission time. A lower transmission time indicates better performance. Let's break down the analysis for each scenario:

**WiFi Network (Bitrate: 1,185,883.50 bps):**

1. Original Files (No Compression):
   - Original Video File 1 (14,929,920 bytes) took 6.367 seconds.
   - Original Video File 2 (253,808,640 bytes) took 35.733 seconds.
2. Bzip2 Compression:
   - File 1 (9,852,909 bytes after compression) took 2.933 seconds.
   - File 2 (166,534,384 bytes after compression) took 21.767 seconds.
3. Proposed Algorithm Compression:
   - File 1 (134,101 bytes after compression) took 1.567 seconds.
   - File 2 (2,332,089 bytes after compression) took 1.967 seconds.

Now, calculating the percentage decrease in transmission time for each compression scenario compared to the original file transmission times on the Wi-Fi network:

Bzip2 Compression:

- For File 1, the transmission time decreased by (6.367 - 2.933) / 6.367 * 100% ≈ 53.85%.
- For File 2, the transmission time decreased by (35.733 - 21.767) / 35.733 * 100% ≈ 39.10%.

Proposed Algorithm Compression:

- For File 1, the transmission time decreased by (6.367 - 1.567) / 6.367 * 100% ≈ 75.30%.
- For File 2, the transmission time decreased by (35.733 - 1.967) / 35.733 * 100% ≈ 94.48%.

**Rating the Performance Improvement:** Our compression technique achieved the biggest percentage reduction in transmission time, notably for larger files, while Bzip2 and ours both markedly improved WiFi network performance. In brief, our compression technique demonstrated remarkable enhancements in performance on GSM/GPRS and WiFi networks, demonstrating its effectiveness in transporting big files across limited networks.

**Compression Speed:** Different algorithms and file types showed different compression speeds in the testing. Both methods performed image compression quickly—between one and two seconds. Our technique achieved effective speed in video compression. Finding the percentage reduction in transmission time for every compression scenario over the WiFi network with relation to the original file:

Bzip2 Compression:

- For File 1, the transmission time decreased by (6.367 - 2.933) / 6.367 * 100% ≈ 53.85%.
- For File 2, the transmission time decreased by (35.733 - 21.767) / 35.733 * 100% ≈ 39.10%.

Proposed Algorithm Compression:

- For File 1, the transmission time decreased by (6.367 - 1.567) / 6.367 * 100% ≈ 75.30%.
- For File 2, the transmission time decreased by (35.733 - 1.967) / 35.733 * 100% ≈ 94.48%.

In conclusion, our suggested compression algorithm performed exceptionally well in GSM and Wi-Fi, which is critical for practical uses. Both algorithms performed image compression quickly, requiring between one and two seconds. Our approach fared better than bzip2 in the video compression task, taking only 5 to 7 seconds to compress a 135 MB video. These speeds confirm that both algorithms are suitable for near-real-time or real-time applications.

**Effectiveness on Different Data Types:** The trials evaluated the performance of the algorithm on both photos and movies. We achieved significant file size reduction with our superior video compression. On the other hand, the general-purpose algorithm bzip2 demonstrated versatility by working equally well for both image and video compression.

**Trade-off Between Compression Ratio and Compression Speed:** Trade-off Between Compression Ratio and Speed: High compression ratios and quick speeds are frequently trade-offs that compression algorithms must make. Although it operates at a little slower pace than bzip2, our video-optimized algorithm delivers exceptional ratios. bzip2, on the other hand, effectively balances speed and compression ratio. Depending on the particular requirements of the application, one of these algorithms—which favor faster speeds or better ratios—should be used.

**Practical Application and Recommendations:** The compression powers of both techniques are impressive. Our technique is well-suited for video-focused applications such as streaming and surveillance, and bzip2's flexibility allows it to be used in scenarios involving a variety of data kinds, including data storage and multimedia content dissemination.

**Transmission and throughput analysis:** Depending on file sizes, 2G and 3G networks' limited bandwidth causes slow transmission, which takes an average of 209 to 2800 seconds. Transmission times for larger files are longer due to network tolerance and latency. The 134,000-byte original movie and bzip2 files encountered transmission issues because of cellular network inherent restrictions. Transmission times can range from 209 seconds for tiny files to 2800 seconds for bigger ones due to bandwidth limitations.  This variation results from the basic idea that larger files take longer to move through the network's constrained pipeline.

Problems including transmission delay tolerance, latency, and timeouts add complexity to data transfer in these networks. When data packets take longer than the allocated time to arrive, the network may get congested, resulting in timeouts. Delays can occur in 2G and 3G networks due to excessive latency, or the amount of time data takes to travel from transmitter to recipient. When delays above the network's tolerance, these variables affect efficiency and dependability and can result in lost connections and unsuccessful transfers. bzip2 and other compression techniques may not be able to resolve issues in certain networks even though they reduce file sizes. Data interchange times are exacerbated by the fact that even big compressed files impede transmission over 2G/3G networks. Using aggressive compression or segmenting larger files are examples of mitigation measures. Reliability during crucial data transfers is improved by error correction, traffic priority, and faster networks like 4G or 5G, which address capacity limitations. Essentially, the constraints of 2G and 3G necessitate careful planning for dependable and effective data transfer.

Lastly, our suggested algorithm's trials with bzip2 demonstrate how effective it is at lowering file sizes while balancing performance and compression ratio trade-offs. With processing rates, compression ratios, and data kinds all taken into account, network administrators can optimize data transmission with the help of these insights. Decision-making in practical data compression applications is aided by the experiments' contribution to our understanding of algorithm effectiveness.

Analysis And Evaluation of Image Compression Algorithm Experiments

Table 3. Shows experiment results for transmitting the image-compressed files using the proposed algorithm in comparison with Bzip2.

| **Camera Pixel Setting** | **Original Size(bytes)** | **Bzip2 Final Size(bytes)** | **Image Algorithm Final Size(bytes)** | **Bzip2 Compression Ratio** | **Image Algorithm Compression Ratio** |
|---|---|---|---|---|---|
| 100 x 100 | 10,305,553 | 6,590,003 | 34,236 | 1.56 | 301.02 |
| 1000 x 1000 | 10,906,757 | 7,231,684 | 605,822 | 1.51 | 18.00 |
| 1500 x 1500 | 11,809,079 | 8,135,437 | 1,435,811 | 1.45 | 8.22 |
| 2000 x 2000 | 12,783,673 | 9,103,772 | 12,783,673 | 1.40 | 1.00 |
| 3000 x 3000 | 13,842,214 | 10,102,988 | 3,563,614 | 1.37 | 3.88 |
| 4000 x 4000 | 15,619,522 | 11,794,307 | 5,449,154 | 1.32 | 2.87 |
| 5000 x 5000 | 17,730,245 | 13,764,439 | 8,064,902 | 1.29 | 2.20 |
| 6000 x 6000 | 20,118,178 | 15,853,431 | 9,996,348 | 1.27 | 2.01 |
| 7000 x 7000 | 22,780,761 | 18,175,600 | 12,616,025 | 1.25 | 1.81 |
| 8000 x 8000 | 25,680,957 | 20,679,516 | 15,431,104 | 1.24 | 1.66 |

To analyze the performance improvement achieved by different compression algorithms in transferring large files across the network, we can calculate the percentage increase in speed for each scenario compared to the original transmission time. Let's break down the analysis for each compression method:

**Original File Sizes:**

- Original image file: 10,000,000 bytes
- Larger file: 25,660,000 bytes

**Bzip2 Compression:**

- Compressed sizes: 6,500,000 bytes (original image) and 20,577,000 bytes (larger file)
- Transmission times: 2.5 seconds (original image) and 4.2 seconds (larger file)

Now, let's calculate the percentage increase in performance for Bzip2 compression:
For the original image file:

$$PercentageIncrease = \frac{(OriginalTime - Bzip2Time)}{(OtiginalTime)} X100$$

$$PercentageIncrease = \frac{(3.3 - 2.5)}{(3.3)} X100$$

For the larger file:

$$PercentageIncrease = \frac{(4.9 - 4.2)}{(4.9)} X100 \approx 14\%$$

Proposed Image Compression Algorithm:

- Compressed sizes: 31,168 bytes (original image) and 15,398,000 bytes (larger file)
- Transmission times: 1.533 seconds (original image) and 4.423 seconds (larger file)

Now, let's calculate the percentage increase in performance for image compression:
For the original image file:

$$PercentageIncrease = \frac{(3.3 - 1.533)}{(3.3)} X 100 \approx 53\%$$

For the larger file:

$$PercentageIncrease = \frac{(4.9 - 4.423)}{(4.9)} X 100 \approx 10\%$$

Performance Ratings:

- Bzip2 achieved approximately a 24% improvement in transmission speed for the original image and a 14% improvement for the larger file.
- Our Proposed algorithm, on the other hand, showed a more substantial improvement, with a 53% increase in performance for the original image and a 10% improvement for the larger file.

In terms of performance improvement, our proposed image algorithm compression outperforms Bzip2 in both cases, particularly for the original image where it achieved a significant speedup. However, it's important to note that the choice between these compression methods should consider factors like the acceptable loss of quality, file type, and specific use case requirements.

Analysis and Evaluation of Aggregation Algorithm for Beehive Data

Experiments evaluated the effectiveness of data aggregation from many sensors, such as vibration, temperature, humidity, and carbon dioxide, in order to optimize the network. Data was collected and stored as a CSV file using a Raspberry Pi, with the goal of assessing transmission times for various volumes of aggregated data. Data aggregation was evaluated with different numbers and sizes of rows, as it is an important technique to reduce the volume of delivered data. Consistent transmission speeds for aggregated data were demonstrated by the results, with a single 46-byte row taking 1.947 seconds, demonstrating the algorithm's efficacy in improving network efficiency.

Table 4. Average time taken to transfer CSV files across the network using data aggregation algorithm.

| **Number of Rows in csv** | **Size(bytes )** | **T1 (s)** | **T2 (s)** | **T3 (s)** | **Average Time taken to send to server(s)** |
|---|---|---|---|---|---|
| 1 | 46 | 1.909 | 1.76 | 2.173 | 1.947 |
| 2 | 90 | 2.106 | 2.072 | 1.945 | 2.041 |
| 3 | 135 | 2.097 | 2.054 | 1.741 | 1.964 |
| 4 | 180 | 1.906 | 1.802 | 1.945 | 1.884 |
| 5 | 226 | 1.816 | 2.146 | 1.801 | 1.921 |
| 6 | 271 | 1.755 | 1.917 | 1.83 | 1.834 |
| 7 | 316 | 1.765 | 1.763 | 2.293 | 1.940 |
| 8 | 361 | 1.935 | 1.838 | 1.831 | 1.868 |
| 9 | 405 | 1.735 | 1.831 | 1.905 | 1.824 |
| 10 | 451 | 2.012 | 1.932 | 1.763 | 1.902 |
| 15 | 676 | 1.932 | 1.858 | 1.902 | 1.897 |
| 30 | 1353 | 1.951 | 1.891 | 1.738 | 1.860 |
| 40 | 1804 | 1.921 | 2.112 | 2.014 | 2.016 |
| 50 | 2255 | 1.796 | 1.904 | 1.692 | 1.797 |
| 100 | 4451 | 1.857 | 1.847 | 1.781 | 1.828 |
| 100000 | 5596285 | 6.177 | 3.912 | 3.496 | 4.528 |

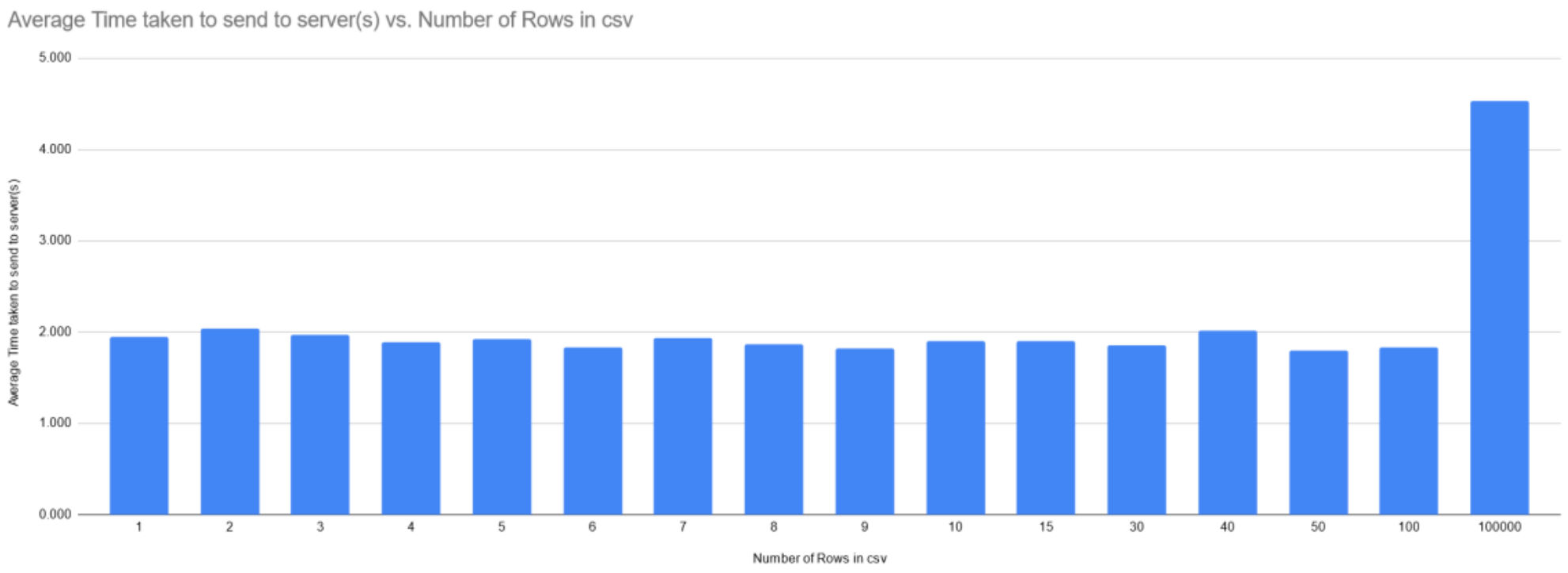

Figure 8: Average Time taken to send to server(s) vs. Number of Rows in CSV

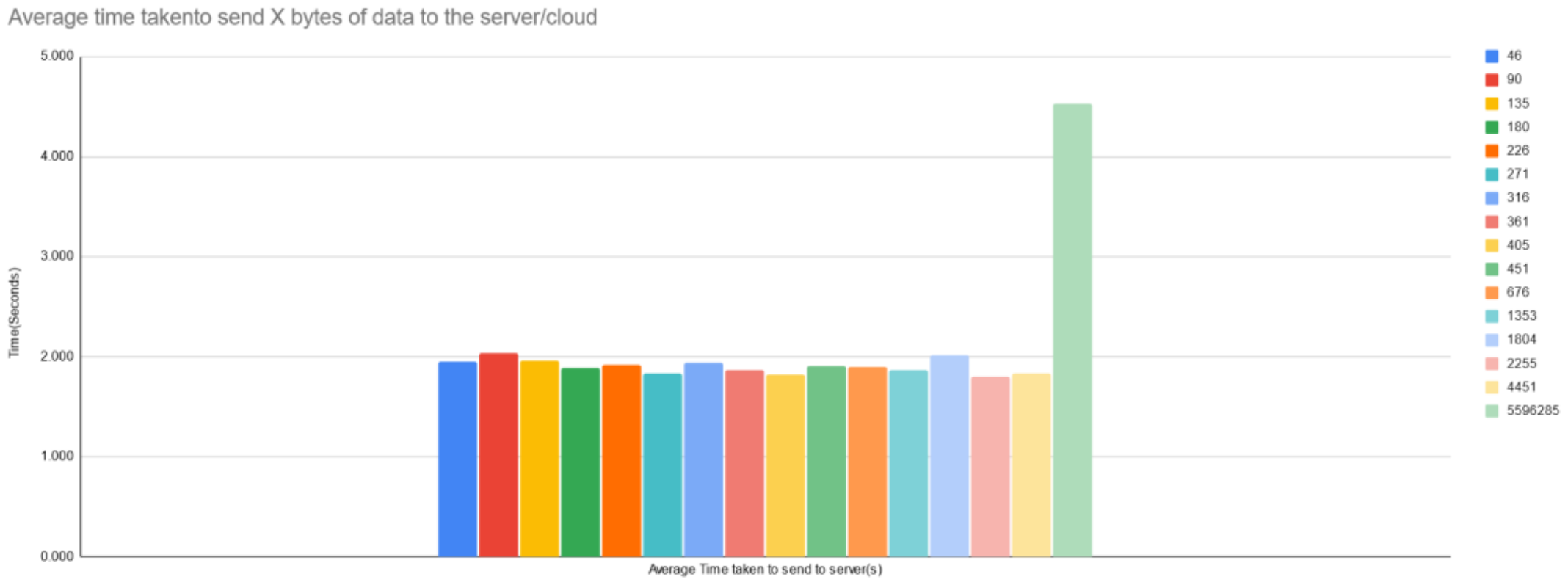

Figure 9. Average time taken to send X bytes of data to the server/cloud

**Efficiency Across Data Quantities:** The aggregation technique demonstrated continuous efficiency, requiring 2.041 seconds for 2 rows (90 bytes), 1.965 seconds for 3 rows (135 bytes), and 1.884 seconds for 4 rows (180 bytes). The increase in transmission time was negligible even at larger sizes, such as 10 rows (451 bytes) in 1.902 seconds, 100 rows (4451 bytes) in 1.828 seconds, and 10000 rows (5596285 bytes) in 4.528 seconds.

**Wi-Fi Impact and Potential Optimization:** Tests limited to Wi-Fi acknowledged the possibility of network unpredictability. Consistent patterns across various data volumes highlight the algorithm's intrinsic effectiveness in streamlining network transmissions, even in the face of Wi-Fi interference.

**Practical Implications and Future Considerations:** The trials verify that the aggregation approach for beehive data transmission optimization is both feasible and effective. Promising are its consistent transmission times when paired with more collected data. Future studies could examine how well it performs in various network scenarios and whether it interacts with other communication protocols like cellular networks or Lo-Ra.

**Conclusion:** In conclusion, the trials effectively demonstrated the potential advantages of the aggregation method for beehive data collection network transmission optimization, offering insightful information for IoT application performance and making a meaningful contribution to the field of network optimization.

## 8. Lo-Ra as a Potential Solution for Remote Locations and Weather Instances

Currently several research studies have been proposed to extend the applicability of our techniques to remote locations and adverse weather conditions. The integration of Lo-Ra technology [8] is suggested to address challenges specific to these environments, emphasizing the need for robust data transfer solutions in diverse agricultural settings.

## 9. Conclusion and Future Research

### 9.1. Conclusion

This research provides a comprehensive solution for data transfer optimization in robust sensor network management specifically using local beehive data files.. In contexts with limited resources, the suggested method for combining edge computing, data compression, and aggregation allows for the real-time gathering, sending, and analysis of massive data files. It has the potential to have a significant influence on beekeeping by boosting honey production, lowering expenses, and enhancing hive health. This method makes it possible to create predictive models, optimize bee foraging depending on environmental conditions, and monitor in real-time. Its usefulness goes beyond beekeeping to include other sensor network applications that encounter difficulties while transmitting massive data files. All things considered, this work makes a substantial contribution to sensor network data optimization, which is advantageous for resource-constrained industries like beekeeping.

### 9.2. Future Research Directions

This research opens up several avenues for future research for example Investigating the application of deep learning for data compression and aggregation in beehive management and other IoT applications is one interesting avenue to pursue. Compared to conventional techniques, deep learning models can achieve even greater compression ratios and more precise data aggregation. Another promising direction is to develop decentralized data processing and storage solutions for beehive management. This would enable beekeepers to process and store data locally, reducing the need for data transmission and improving data security. Finally, it is important to evaluate the proposed approach in real-world application settings. This would involve deploying the system in beehive farms and collecting feedback from beekeepers on its performance and usability.

## 10. ACKNOWLEDGMENTS

Expressions of gratitude are extended to the AdEMNEA Project family, the project leaders and supervisors (Dr. Julianne Sansa Otim, Dr. Mary Nsabagwa, Prof. Marco Zennaro, and Prof. Stephen Wolthusen) that have contributed to the success of this research. This includes acknowledgment of funding sources from NORAD, technical support, and valuable insights provided by collaborators and mentors.

## AUTHORS

**Ephrance Eunice Namugenyi**: PhD Student Data Communications and Software Engineering, Makerere University Uganda, Lecture Department of Electrical and Electronics Engineering, Kyambogo University, Researcher in Computer Networks and Communication Technologies, Business Woman.

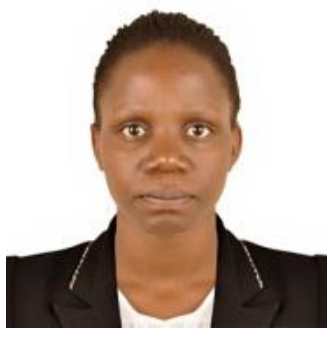

**David Tugume** is an aspiring researcher in IOT who did a Bsc Software engineering at undergrad and now doing a MSc in data communication and software engineering at makerere university Department of networks. He is doing research in detecting faults on LoRa Network.

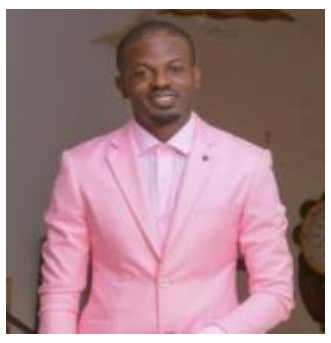

**Augustine Kigwana**: I am a committed student studying software engineering at Makerere University in Uganda. I am very involved in the cutting edge fields of artificial intelligence, embedded systems, and IoT (Internet of Things). I'm passionate in developing technology that can learn, adapt, and get better with time. I hope to use intelligent infrastructure to help progress smart home technologies, healthcare wearables, and urban development.

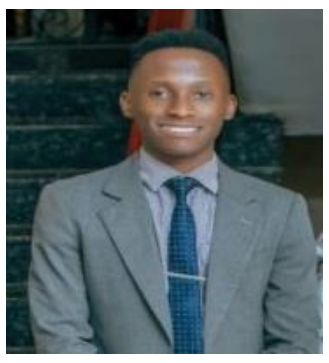

**Benjamin Rukundo** is a motivated student pursuing a bachelor's degree in software engineering at Makerere University with a concentration on artificial intelligence and the Internet of Things, notably natural language processing. He has been actively working on cutting-edge IoT projects lately, creating new channels of communication for edge devices to send sensor data that go beyond Wi-Fi.

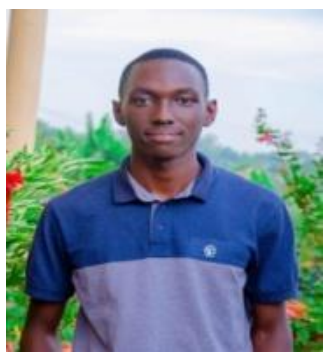